\documentstyle[epsfig,here,12pt,color]{elsart}

\newcommand{\be}[1]{\begin{equation} \label{(#1)}}
\newcommand{\ee}{\end{equation}}
\newcommand{\baq}[1]{\begin{eqnarray} \label{(#1)}}
\newcommand{\eaq}{\end{eqnarray}}
\newcommand{\rf}[1]{(\ref{(#1)})}
\newcommand{\ba}{\begin{array}}
\newcommand{\ea}{\end{array}}

\newcommand{\CH}{\tilde\chi}

\def\beq   {\begin{equation}}
\def\eeq   {\end{equation}}
\def\beqd  {\begin{displaymath}}
\def\eeqd  {\end{displaymath}}
\def\beqaa {\begin{eqnarray}}
\def\eeqaa {\end{eqnarray}}

\def\ti  {\tilde}

\def\sz{\ifmmode{\tilde{\chi}^0} \else{$\tilde{\chi}^0$} \fi}
\def\sw{\ifmmode{\tilde{\chi}} \else{$\tilde{\chi}$} \fi}

\newcommand{\lsim}{\;\raisebox{-0.9ex}{$\textstyle\stackrel{\textstyle<}
           {\sim}$}\;}

\def\bold#1{\setbox0=\hbox{$#1$} 
     \kern-.025em\copy0\kern-\wd0 
     \kern.05em\copy0\kern-\wd0 
     \kern-.025em\raise.0433em\box0 }

\begin{document}

\begin{flushright}
   UWThPh-2008-03
\\ IFIC/08-12
 \\BONN-TH-2008-02
\end{flushright}

\begin{frontmatter}
  
  \title{CP observables with spin-spin correlations 
in chargino production}

\author{A.~Bartl$^{a,b}$},
\author{K.~Hohenwarter-Sodek$^a$},
\author{T.~Kernreiter$^a$},
\author{O.~Kittel$^c$} and
\author{M.~Terwort$^d$}

\address{(a) Faculty of Physics, University of Vienna\\ 
Boltzmanngasse 5, A-1090 Wien, Austria}
\address{(b) AHEP Group, Institut de F\'\i sica Corpuscular -- C.S.I.C. \\
Universitat de Val\`encia, Edifici Instituts d'Investigaci\'o \\ 
Apt. 22085, E-46071 Val\`encia, Spain}
\address{
(c) Physikalisches Institut der Universit\"at Bonn\\
Nu{\ss}allee 12, D-53115 Bonn, Germany}
\address{
(d) Institut f\"ur Experimentalphysik, Universit\"at Hamburg\\
Notkestra\ss e 85, D-22607 Hamburg, Germany}
%


\noindent

\begin{abstract}
 We study the CP-violating terms of the spin-spin correlations in chargino 
 production $e^+e^-\to \CH^\pm_1\CH^\mp_2$, and their
 subsequent two-body decays into sneutrinos plus leptons. 
 We propose novel CP-sensitive observables 
 with the help of T-odd products of the spin-spin terms.
 These terms depend on the polarizations of both charginos,
 with one polarization perpendicular to the production plane. 
 We identify two classes of CP-sensitive observables; 
 one requires the reconstruction of the production plane, the other not. 
 Our framework is the Minimal Supersymmetric 
 Standard Model with complex parameters. 
 \end{abstract}
\end{frontmatter}


\section{Introduction}

Supersymmetric (SUSY) extensions of the Standard Model (SM), 
like the Minimal Supersymmetric Standard Model (MSSM)~\cite{mssm},
give rise to new sources of CP violation~\cite{Haber:1997if}. 
From a mathematical 
point of view, this means that in the SUSY Lagrangian 
complex parameters enter whose phases cannot be removed by
redefining the fields. The presence of CP phases
can drastically alter the phenomenology of the underlying model 
(for a recent review, see~\cite{Ibrahim:2007fb}).
For instance, contributions of SUSY CP phases to the electric dipole moments (EDM)
of electron, neutron, and that of the atoms $^{199}$Hg and $^{205}$Tl can
be close or beyond the present experimental upper bounds~\cite{EDMs,EDMs1}, 
and thus in turn constrain the size of these CP phases~\cite{EDMs,EDMs1}.
These constraints, however, are strongly model dependent, see e.g.~\cite{EDMs1}. 
Thus measurements of CP observables outside the EDM sector
are necessary to independently determine or constrain the CP phases. 
Furthermore, non-vanishing phases can significantly change 
masses, cross sections and decay branching ratios
of SUSY particles, compared to the real case, see e.g.~\cite{Brhlik:1998gu,staplex}. 
Hence, in determining the underlying 
model parameters, the effect of their CP phases has to be taken into account.
The phases could be measured once supersymmetric 
particles are accessible at future colliders. 
A genuine signal for CP violation would be 
the measurement of non-vanishing CP-sensitive observables.

In this paper, we propose CP-sensitive observables
in chargino production
\be{eq:production}
e^+e^-\to \CH^\pm_1\CH^\mp_2~,
\ee
within the MSSM. For the processes $e^+e^-\to \CH^+_1\CH^-_1,\CH^+_2\CH^-_2$
the CP-sensitive terms in the amplitude squared vanish
at tree level since all coupling factors are real~\cite{TPchargino,Bartl:2004vi}. 
The chargino mass matrix, in the weak basis, is given by
\be{eq:Charmass}
\mathcal{M}_C=\left(\begin{array}{ccc}
M_2 & \sqrt{2}m_W \sin\beta\\[2mm]
\sqrt{2}m_W \cos\beta & \mu
\end{array}\right),
\ee
with $M_2$ the SU(2) gaugino mass parameter, and $\tan\beta$ the 
ratio of the vacuum expectation values of the two Higgs fields.
$M_2$ and $\tan\beta$ can be chosen real and positive, while
the higgsino mass parameter 
can be complex  $\mu=|\mu| e^{i\phi_\mu}$.
At tree level, $\phi_\mu$ is the only CP-violating phase 
that gives rise to CP-sensitive observables in chargino production,
while phases from other sectors can contribute at loop
level~\cite{Rolbiecki:2007se,Osland:2007xw}. 
For example, phases of the gaugino mass parameter $M_1$, and
the trilinear coupling parameters $A_{t}$ in the stop-sector,
lead to rate differences in $\CH^+_1\CH^-_2$ production and that 
of the charge conjugated pair $\CH^-_1\CH^+_2$
at the percent level~\cite{Rolbiecki:2007se}.
For chargino decays, 
rate asymmetries of the partial chargino decay widths
can exceed $10\%$, mainly due to the phases
of $M_1$ and $A_{t,b}$~\cite{Yang:2002am} (see also \cite{Fujimoto:2007bn}).

Another class of promising CP-sensitive observables are based
on so-called {\it T-odd correlations} (or {\it T-odd products}),
see e.g.~\cite{Valencia:1994zi}. 
They can give rise to CP-violating effects already at tree-level, and therefore 
suffer not from loop suppression as rate asymmetries.
Previous studies of CP-sensitive observables, based on T-odd
products in chargino production and decay, have been focussing on 
the spin correlations between production and decay of only 
one chargino~\cite{TPchargino,Bartl:2004vi}.
The corresponding terms in the amplitude squared 
involve the polarization vector perpendicular to the
production plane of one of the produced charginos.
Such a  transverse polarization component is a genuine signal of CP violation.
The transverse polarization is then retrieved from asymmetries
in the azimuthal distribution of the decay products,
which can be as large as $30\%$, even for small 
$\phi_\mu$ of order $\pi/10$~\cite{Bartl:2004vi}.
In such an analysis of the spin-correlations,
the polarization, i.e.
the decay, of only one chargino needs to be considered. 
However, if the decays of both charginos are taken into
account, one can probe their spin-spin correlations~\cite{MoortgatPick:1998sk}.
These are terms in the amplitude squared that include the 
polarization vectors of both charginos.
The angular distributions of the decay products of the two charginos
are correlated to one another due to total angular momentum conservation.
Spin-spin correlations in chargino production
and decay have been utilized for the determination
of CP-even coupling factors~\cite{MoortgatPick:1998sk,Choi:1998ut}.
Moreover, spin-spin correlations have been used for the definition
of CP-sensitive observables in the decays of third generation 
squarks~\cite{CPstop,CPsbottom}.

In the present paper, we propose novel CP-sensitive
observables with the help of T-odd products in the chargino spin-spin 
correlation terms. We take the decay of both charginos into account,
and consider, for definiteness, their subsequent leptonic two-body decays  
\be{eq:decay}
\CH^+_i\to \tilde\nu_\ell ~\ell^+~, \quad
\CH^-_j\to\bar{\tilde\nu}_{\ell^\prime} ~\ell^{\prime-}~,
\quad i,j=1,2~(i\neq j)~,
\quad \ell,\ell^{\prime }=e,\mu~.
\ee
By analyzing the CP-odd parts of the spin-spin terms, we find 
two independent T-odd products.
The first one includes the momenta of the beams and those of 
the two decay leptons.
Thus a reconstruction of the production plane is not necessary 
for a measurement of the corresponding CP-sensitive observables.
From an experimental perspective, this seems to be advantageous compared
to CP-sensitive observables in chargino production and leptonic decays, 
where such a reconstruction is essentially required~\cite{TPchargino,Bartl:2004vi}.
The second T-odd product which we find
involves the momenta of the charginos, requiring the reconstruction
of the production plane.
We consider two sorts of CP-sensitive observables 
which we obtain from the spin-spin correlations. Defining 
their statistical significances, we can make a quantitative
comparison of their accessibility.

The paper is organized as follows.
In Section~\ref{cross section}, we present analytical
formulae for the amplitude squared of chargino production and decay
$e^+e^-\to \CH^+_i\CH^-_j\to 
\tilde\nu_\ell~\ell^+~\bar{\tilde\nu}_{\ell'}~\ell'^-$.
We identify the T-odd products that are involved
in the spin-spin terms of the amplitude squared in Section~\ref{identify}, 
and define the associated CP-sensitive observables in 
Section~\ref{observables}.
We present numerical results in Section~\ref{numerics}.
In Section~\ref{conclusion}, we give a summary and conclusions.

\section{Cross section \label{cross section}}

\begin{figure}
\hspace{1cm}
\begin{minipage}[t]{3.5cm}
\begin{center}
{\setlength{\unitlength}{1cm}
\begin{picture}(1.5,2.5)
\put(-2.,-1.1){\includegraphics{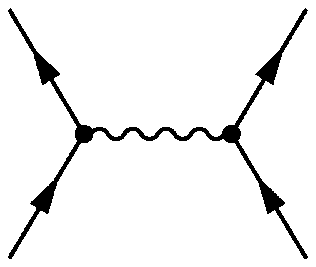}}
\put(-2.0,-1.6){$e^{-}$}
\put(1.4,-1.6){$\tilde\chi^-_j$}
\put(-2.0,1.5){$e^{+}$}
\put(1.4,1.5){$\tilde\chi^+_i$}
\put(-.3,.4){$\gamma$}
\end{picture}}
\end{center}
\end{minipage}
\hspace{2cm}
\vspace{.8cm}

\begin{minipage}[t]{3.5cm}
\begin{center}
{\setlength{\unitlength}{1cm}
\begin{picture}(-2.5,2.5)
\put(3.,2.3){\includegraphics{prog.ps}}
\put(3.,1.7){$e^{-}$}
\put(6.4,4.9){$\tilde\chi^+_i$}
\put(3.,4.8){$e^{+}$}
\put(6.4,1.8){$\tilde\chi^-_j$}
\put(4.7,3.7){$Z$}
\end{picture}}
\end{center}
\end{minipage}
\hspace{2cm}
\vspace{.8cm}

\begin{minipage}[t]{3.5cm}
\begin{center}
{\setlength{\unitlength}{1cm}
\begin{picture}(2.5,2.5)
\put(11.8,5.4){\includegraphics{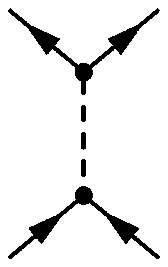}}
\put(11.5,8.2){$e^{+}$}
\put(11.5,5.1){$e^{-}$}
\put(13.4,8.2){$\tilde\chi^+_i$}
\put(13.4,5.0){$\tilde\chi^-_j$}
\put(12.1,6.6){$\tilde{\nu}$}
 \end{picture}}
\end{center}
\end{minipage}
\vspace{-4.6cm}
\caption{\label{Fig1} 
Feynman diagrams for chargino production 
$e^+e^-\rightarrow\tilde\chi^+_i\tilde\chi^-_j$~\cite{MoortgatPick:1998sk}.}
\end{figure}

Chargino production 
$e^+e^-\to \tilde\chi^+_i\tilde\chi^-_j$
proceeds via $\gamma, Z$ exchange in the $s$-channel,
and $\tilde\nu$ exchange in the $t$-channel, see the
Feynman diagrams in Fig.~\ref{Fig1}.
The $\gamma$ exchange vanishes for non-diagonal
chargino production $e^+e^-\to \tilde\chi^\pm_1\tilde\chi^\mp_2$.
For diagonal chargino production,
$e^+e^-\to \tilde\chi^+_i\tilde\chi^-_i$, the squared amplitude
contains no CP-violating terms at tree level~\cite{TPchargino,Bartl:2004vi}.

We write the differential cross section for chargino production and decay
$e^+e^-\to \CH^+_i\CH^-_j\to \tilde\nu_\ell ~\ell^+ \bar{\tilde\nu}_{\ell'}
\ell'^-$, $\ell,\ell'=e,\mu$, generically as
\be{eq:crossection}
{\rm d}\sigma=\frac{1}{2~s}~|T|^2~{\rm dLips}~,
\ee
with the center-of-mass energy $\sqrt s$,  
and the Lorentz invariant phase space element ${\rm dLips}$, which 
can be found in \cite{Bartl:2004vi}.
The amplitude squared $|T|^2$ was calculated in 
Ref.~\cite{MoortgatPick:1998sk}
in the spin density matrix
formalism\footnote[1]{For a detailed discussion of the 
spin density matrix formalism, we refer to Ref.~\cite{Haber:1994pe}.}
\baq{eq:amplitudesquared}
|T|^2&=&4|\Delta(\CH_i)|^2|\Delta(\CH_j)|^2
\left[
P~D_i~D_j+
\sum^3_{a=1}\Sigma^a_P~\Sigma^a_{D_i}~D_j\right.\nonumber\\[2mm]
{}&&\left.+\sum^3_{b=1}\Sigma^b_P~\Sigma^b_{D_j}~D_i
+\sum^3_{a,b=1}\Sigma^{ab}_P~\Sigma^a_{D_i}~\Sigma^b_{D_j}
\right]~,
\eaq
with the propagators $\Delta(\CH_{i,j})=1/[p^2_{\chi_{i,j}}-m^2_{\chi_{i,j}}
+im_{\chi_{i,j}}\Gamma_{\chi_{i,j}}]$ of the decaying charginos.
The amplitude squared has contributions
from chargino production ($P$) and decay ($D$).
The terms $P$ and $D_i$, $D_j$ are those parts of the
spin density production and decay matrices, respectively,
that are independent of the polarizations of the charginos.
The contributions $\Sigma^a_P$ and
$\Sigma^a_{D_i}$ depend on the polarization
basis vectors $s_{\chi^+_i}^a$ (for their definition 
see Appendix~\ref{Vectors}, Eq.~\rf{eq:polvec}) of the decaying 
chargino $\tilde{\chi}^+_i$, while $\Sigma^b_P$ and $\Sigma^b_{D_j}$ 
depend on the polarization basis vectors $s_{\chi^-_j}^b$ of the 
decaying chargino $\tilde{\chi}^-_j$.
We choose a coordinate frame such that $a,b =3$ denote the
longitudinal polarizations, $a,b =1$ the transversal
polarizations in the production plane, and $a,b =2$ the
transversal polarizations perpendicular to the production plane.
The quantities $D_i$, $D_j$, $\Sigma^a_{D_i}$ and $\Sigma^b_{D_j}$ are given in
Appendix~\ref{Decaydensity}.
The full expressions for the quantities 
$P$, $\Sigma^a_P$, $\Sigma^b_P$ and
$\Sigma^{ab}_P$ can be found in Ref.~\cite{MoortgatPick:1998sk}. 

\medskip

The contributions to the amplitude squared
which depend on the polarizations of both charginos are the spin-spin 
correlation terms $\Sigma^{ab}_P$.
The T-odd parts of the spin-spin correlation terms 
are from pure $Z$ exchange $(Z Z)$ and from  $Z$-$\tilde\nu$ 
interference $(Z \ti\nu_e)$, and are those
which include one chargino spin vector with a component 
perpendicular to the production plane, i.e., those with 
$ab=12,21,23,32$~\cite{MoortgatPick:1998sk}
\be{eq:Toddpart1}
\Sigma^{ab}_P(Z Z)
=\frac{g^4}{\cos^4\Theta_W}|\Delta(Z)|^2 (L^2_e c_{-+}+R^2_e c_{+-})~
{\rm Im}(O_{ij}^{'L} O_{ij}^{'R*})~ f^{ab}~,
\ee
\baq{eq:Toddpart2}
\Sigma^{ab}_P(Z \ti\nu_e)
=-\frac{g^4}{2\cos^2\Theta_W}
\Delta(Z)\Delta(\ti\nu_e)^*L_e~c_{-+}~
{\rm Im}(V_{i1}^* V_{j1} O_{ij}^{'R})~ f^{ab}\,,
\eaq
where the left and right chiral couplings of the charginos to the $Z$ boson
are 
\be{eq:OLp}
O'^L_{ij}=-V_{i1}V^*_{j1}-\frac{1}{2}V_{i2}V^*_{j2}+
\delta_{ij}\sin^2\Theta_W~,
\ee
\be{eq:ORp}
O'^R_{ij}=-U^*_{i1}U_{j1}-\frac{1}{2}U^*_{i2}U_{j2}+
\delta_{ij}\sin^2\Theta_W~,
\ee
with $\Theta_W$ the weak mixing angle and the unitary 
$2\times2$ mixing matrices $U$ and $V$ 
which diagonalize the chargino mass matrix, Eq.~\rf{eq:Charmass}, 
$U^{\ast} {\mathcal M_C} V^{-1}={\rm diag}(m_{\chi_1},m_{\chi_2})$.
In Eqs.~\rf{eq:Toddpart1} and~\rf{eq:Toddpart2}, $g$ is the
SU(2) weak coupling constant,
$L_e=-1/2+\sin^2\Theta_W$, $R_e=\sin^2\Theta_W$, $\Delta(Z)=i/(s-m^2_Z)$,
$\Delta(\ti\nu_e)=i/(t-m^2_{\ti\nu_e})$, with $s=(p_{e^-}+p_{e^+})^2$,
$t=(p_{e^-}-p_{\chi^-_j})^2$, and $m_{\ti\nu_e}$ ($m_Z$) is the mass of the
electron sneutrino ($Z$ boson).
The dependence on the beam polarizations is given by the factors
\baq{eq:beampolfac}
c_{+-}&=&(1+{\mathcal P}_-)(1-{\mathcal P}_+)~,\nonumber\\[2mm]
c_{-+}&=&(1-{\mathcal P}_-)(1+{\mathcal P}_+)~,
\eaq
where ${\mathcal P}_-$ and ${\mathcal P}_+$ are the
degrees of longitudinal polarization of the electron and positron beam, 
respectively, with $-1\leq{\mathcal P}_\pm \leq 1$.
The kinematical dependence of the spin-spin correlation terms,
Eqs.~\rf{eq:Toddpart1} and~\rf{eq:Toddpart2}, is given  by the 
function~\cite{MoortgatPick:1998sk}
\baq{eq:f8}
f^{ab}&=&~\varepsilon_{\mu\nu\rho\sigma}~
\Big[s_{\chi^-_j}^{b,\,\mu}~ s_{\chi^+_i}^{a,\,\nu}~ p_{\chi^+_i}^\rho~p_{e^-}^\sigma~
(p_{e^+}\!\cdot\! p_{\chi^-_j})
-s_{\chi^-_j}^{b,\,\mu}~ s_{\chi^+_i}^{a,\,\nu}~ p_{e^+}^\rho~p_{\chi^-_j}^\sigma~
(p_{\chi^+_i}\!\cdot\! p_{e^-})\nonumber\\
{}&&
 +s_{\chi^+_i}^{a,\,\mu}~ p_{\chi^+_i}^\nu~ p_{e^-}^\rho~ 
p_{\chi^-_j}^\sigma~(p_{e^+}\!\cdot\! s_{\chi^-_j}^{b})
 +s_{\chi^-_j}^{b,\,\mu}~ p_{\chi^+_i}^\nu~p_{e^+}^\rho~p_{\chi^-_j}^\sigma~
(p_{e^-} \!\cdot\!s_{\chi^+_i}^a)\Big]~,
\eaq
with $\varepsilon_{0123}=-1$.

Note that the spin-spin correlation terms in Eqs.~\rf{eq:Toddpart1}
and~\rf{eq:Toddpart2} depend 
on the imaginary parts of the products of chargino couplings,
${\rm Im}(O_{ij}^{'L} O_{ij}^{'R*})$ and
${\rm Im}(V_{i1}^* V_{j1} O_{ij}^{'R})$, and thus are
manifestly CP-sensitive, i.e., sensitive to the
phase $\phi_\mu$ of the chargino sector.
We also give the spin-spin correlation terms in the
laboratory system in Appendix~\ref{Spin-spin-terms}.

\section{Identifying the T-odd products in the 
         spin-spin correlation terms \label{identify}}

For an identification of the T-odd products in chargino production
and decay, we consider 
the kinematical dependence of the spin-spin correlation terms
of the amplitude squared, Eq.~\rf{eq:amplitudesquared}, 
\baq{eq:kinedep}
\sum^3_{a,b=1}\Sigma^{ab}_P~\Sigma^a_{D_i}~\Sigma^b_{D_j} \propto
\sum^3_{a,b=1}f^{ab}\cdot (s^{a}_{\chi^+_i} \!\cdot\! p_{\ell^+})
\cdot (s^{b}_{\chi^-_j} \!\cdot\! p_{\ell'^-}) =
\nonumber\\[3mm]
\epsilon_{\mu\nu\rho\sigma}\Big[
p_{\ell'^-}^\mu p_{\ell^+}^\nu p_{\chi^+_i}^\rho p_{e^-}^\sigma~
(p_{e^+} \!\cdot\! p_{\chi^-_j})
+ p_{\ell'^-}^\mu p_{\ell^+}^\nu p_{\chi^-_j}^\rho p_{e^+}^\sigma~
(p_{e^-}  \!\cdot\!p_{\chi^+_i})\nonumber\\[3mm]
+ p_{\ell^+}^\mu p_{\chi^+_i}^\nu p_{e^-}^\rho p_{\chi^-_j}^\sigma~
(p_{e^+}  \!\cdot\!p_{\ell'^-})
+ p_{\ell'^-}^\mu p_{\chi^+_i}^\nu p_{e^+}^\rho p_{\chi^-_j}^\sigma~
(p_{e^-}  \!\cdot\!p_{\ell^+})\Big]~,
\eaq
where the scalar products $(s^{a}_{\chi^+_i}\!\cdot p_{\ell^+})$ and  
$(s^{b}_{\chi^-_j}\!\cdot  p_{\ell'^-})$
appear in $\Sigma^a_{D_i}$ and $\Sigma^b_{D_j}$, respectively,
see Eq.~\rf{eq:Slep}.
We have used the explicit expression for $f^{ab}$, Eq.~\rf{eq:f8},
and the completeness relation for the chargino spin 
vectors~\cite{MoortgatPick:1998sk,Haber:1994pe} 
\baq{eq:chicompleteness}
\sum_c s_{\chi^\pm_k}^{c,\,\mu} \cdot s_{\chi^\pm_k}^{c,\,\nu}=
-g^{\mu\nu}+\frac{p_{\chi^\pm_k}^\mu p_{\chi^\pm_k}^\nu}{ m_{\chi_k}^2}~. 
\eaq
If we now substitute in the center-of-mass system 
the chargino 3-momenta by the corresponding 
lepton 3-momenta $\vec{p}_{\chi^+_i}\to \vec{p}_{\ell^+}$,  
$\vec{p}_{\chi^-_j}\to \vec{p}_{\ell'^-}$,
on the right hand side of Eq.~\rf{eq:kinedep},
we find the T-odd product
\be{eq:Todd1}
{\mathcal O}_T=
\hat{p}_{e^-}\cdot(\hat{p}_{\ell^+}+\hat{p}_{\ell'^-})~
\hat{p}_{e^-}\cdot(\hat{p}_{\ell^+}\times \hat{p}_{\ell'^-})~,
\ee
of the unit momentum vectors 
$\hat p = \vec p/ |\vec p|$.
If we do not replace the chargino momenta, we find  
an additional T-odd product
\baq{eq:Todd1prod}
{\mathcal O}^{\rm prod}_T&&=(\hat p_{e^-}\cdot \hat p_{\ell'^-}) ~
 \hat p_{e^-}\cdot(\hat p_{\chi^-_j}\times\hat p_{\ell^+})+
(\hat p_{e^-}\cdot \hat p_{\ell^+})~
 \hat p_{e^-}\cdot(\hat p_{\chi^-_j}\times\hat p_{\ell'^-})~.
\eaq
Since the T-odd product ${\mathcal O}^{\rm prod}_T$ includes the chargino 
momentum $\hat p_{\chi^-_j}$, it will require the reconstruction 
of the production plane.
However, as we will see later, this product 
gives rise to larger CP-sensitive observables.
Note that in writing all momentum vectors as unit vectors, 
the T-odd products
${\mathcal O}_T$ and ${\mathcal O}^{\rm prod}_T$ are dimensionless.

\section{CP-sensitive observables \label{observables}}

In this Section, we 
define our CP-sensitive observables, which depend on the 
T-odd parts of the spin-spin correlations for  
chargino production and decay.
For an operator ${\mathcal {\widehat O}}$, we define its expectation value by
\be{eq:Obs1}
\langle {\mathcal {\widehat O}} \rangle=
\frac{ \int{\mathcal {\widehat O}}~ |T|^2~ {\rm dLips} } 
     {\int |T|^2~ {\rm dLips} }
=
\frac{1}{\sigma} \int~{\mathcal {\widehat O}}~ 
\frac{{\rm d}\sigma}{\rm dLips}~{\rm dLips}~.
\ee
We now define two classes of CP observables;
one class requires the reconstruction of the chargino momenta,
the other class not.

\subsection{CP-sensitive observables without the knowledge 
            of the production plane\label{cp1}}

Using Eq.~\rf{eq:Obs1} and the T-odd product ${\mathcal O}_T$,
Eq.~\rf{eq:Todd1}, we define the two CP-sensitive observables
\be{eq:Obs}
\langle {\mathcal O}_T\rangle  \qquad {\rm and} \qquad
\langle {\rm Sgn}({\mathcal O}_T)\rangle ~.
\ee
The integration in Eq.~\rf{eq:Obs1} with ${\mathcal O}_T$ and ${\rm Sgn}({\mathcal
  O}_T)$ as operators, projects out the CP-sensitive parts in the spin-spin
correlation terms of the amplitude squared.
The contributions from the terms of the spin correlations between production
and decay in Eq.~\rf{eq:amplitudesquared}, $\Sigma^a_P~\Sigma^a_{D_i}~D_j$ and
$\Sigma^b_P~\Sigma^b_{D_j}~D_i$, cancel each other.
The observable $\langle {\rm Sgn}({\mathcal O}_T)\rangle$ represents an up-down
asymmetry, which gives the relative number of events for which the sign 
of the T-odd product ${\mathcal O}_T$ in Eq.~\rf{eq:Todd1} 
is positive ($N_+/N$), subtracted from the
relative number of events where it is negative ($N_-/N$).
$\langle {\mathcal O}_T\rangle$, on the other hand, gives the expectation value 
of the momentum configuration ${\mathcal O}_T$, Eq.~\rf{eq:Todd1}, for the event sample.
Note that, since ${\mathcal O}_T$ does not include the chargino momenta, the CP-sensitive
observables in Eq.~\rf{eq:Obs} can be probed without the knowledge of 
the production plane.
Note further, that for the production of the charge conjugated 
pair of charginos, $e^+e^-\to \tilde\chi^-_i\tilde\chi^+_j\to 
\bar{\tilde\nu}_\ell ~\ell^- \tilde\nu_{\ell'}\ell'^+$,
the observables  
$\langle {\mathcal O}_T\rangle$
and
$\langle {\rm Sgn}({\mathcal O}_T)\rangle$
change sign. 
In order that the two observables from the two chargino production
processes
$e^+e^-\to \tilde\chi^-_1\tilde\chi^+_2$ and
$e^+e^-\to \tilde\chi^-_2\tilde\chi^+_1$,
do not sum up to zero, one has to distinguish
from which chargino ($\tilde\chi^\pm_1$ or $\tilde\chi^\mp_2$)
the final state leptons originate.
This can be achieved by using the different energy distributions
of the leptons, since their kinematical 
limits depend on the mass of the decaying chargino.
In Subsection~\ref{backgrounds}, we give a numerical example
showing how the leptons can be distinguished by using their 
different energy distributions.

\subsection{CP-sensitive observables which require 
            the knowledge of the production plane\label{cp2}}

For the T-odd product ${\mathcal O}^{\rm prod}_T$, Eq.~\rf{eq:Todd1prod},
we analogously obtain the CP observables
\be{eq:ObsProd}
\langle {\mathcal O}_T^{\rm prod}\rangle  \qquad {\rm and} \qquad
\langle {\rm Sgn}({\mathcal O}_T^{\rm prod})\rangle ~.
\ee
The CP-sensitive observables in Eq.~\rf{eq:ObsProd} not only receive
contributions from the spin-spin correlation terms,
but also from the spin correlation terms between production and decay,
$\Sigma^a_P~\Sigma^a_{D_i}~D_j$ and
$\Sigma^b_P~\Sigma^b_{D_j}~D_i$, see Eq.~\rf{eq:amplitudesquared}.

Since the T-odd product ${\mathcal O}^{\rm prod}_T$ includes the 
chargino momenta, a measurement of the CP-sensitive observables 
requires the reconstruction
of the production plane. The extent to which such a reconstruction
can be accomplished depends on the decay pattern of the produced charginos.
For their subsequent two-body decays which we consider here,
$e^+e^-\to \CH^\pm_1\CH^\mp_2\to \tilde\nu_\ell\ell^+ \bar{\tilde\nu}_{\ell'} \ell'^-$,   
the chargino momentum three-vector can be reconstructed up to a sign
ambiguity in its second component, if the masses of the
involved particles are known~\cite{Bartl:2005uh,Buckley:2007th,ChoiRecon}.
The two solutions (true and false) can then be combined using statistical 
methods, as shown in~\cite{Buckley:2007th}.

\subsection{Theoretical statistical significances\label{statistics}}

We have defined two kinds of CP-sensitive observables, 
$\langle {\mathcal O}\rangle$ and 
$\langle {\rm Sgn}({\mathcal O})\rangle$, 
based on the T-odd products 
${\mathcal O}={\mathcal O}_T, {\mathcal O}_T^{\rm prod}$.  
%
The observable $\langle {\mathcal  O}\rangle$ is obtained  
by matching the complete kinematical (angular) dependence of the spin-spin 
correlation terms in the amplitude squared.
In the literature, this technique is known by the name
\emph{optimal observables}~\cite{ChoiRecon,optimal}.
In order to compare the two kinds of observables,
we define their theoretical statistical significances.
A comparison of the numerical values of $\langle {\mathcal O}\rangle$ and 
$\langle {\rm Sgn}({\mathcal O})\rangle$ alone cannot be used
to decide which observable is more sensitive to the CP phases.
The statistical significances also allow us to compare the observables
which are based on the T-odd product ${\mathcal O}_T$, with those which
are based on ${\mathcal O}_T^{\rm prod}$, which includes the chargino momentum.

The theoretical statistical significance of the CP observable 
$\langle {\mathcal {\widehat O}}\rangle$, where
${\mathcal {\widehat O}}={\mathcal O}_T, {\mathcal O}_T^{\rm prod}$ or
${\mathcal {\widehat O}}={\rm Sgn}({\mathcal O}_T), {\rm Sgn}({\mathcal O}_T^{\rm prod})$,
is defined by~\cite{ChoiRecon,Bartl:2006bn}
\be{eq:EffAsy}
 S[{\mathcal {\widehat O}}]=\sqrt{N}~
\frac{|\langle {\mathcal {\widehat O}}\rangle|}{\sqrt{\langle 
{\mathcal {\widehat O}}^2\rangle}}~,
\ee
with the number of events 
 $N=8~\sigma(e^+e^-\to \tilde\chi^+_1\tilde\chi^-_2)
\times {\rm BR}(\tilde\chi^+_1\to \tilde\nu_e~e^+)
\times {\rm BR}(\tilde\chi^-_2\to \bar{\tilde\nu}_e~e^-)
\times {\mathcal L}$,
where ${\mathcal L}$ denotes the integrated luminosity.
The factor $8$ appears since there are
$4$ possibilities to sum over the lepton flavors $\ell = e, \mu$,
and two charge assignments for chargino production 
$\sigma(e^+e^-\to \tilde\chi^\pm_1\tilde\chi^\mp_2)$.
The quantity $ S[{\mathcal {\widehat O}}]/\sqrt{\mathcal L}$ is called
{\it effective asymmetry} in~\cite{ChoiRecon}.

We have obtained the significance $S[{\mathcal {\widehat O}}]$, Eq.~\rf{eq:EffAsy}, by 
imposing that the observable should at least be 
larger than its absolute statistical error~\cite{Bartl:2006bn}
\be{eq:Error}
\frac{|\langle {\mathcal {\widehat O}}\rangle|}{\Delta\langle {\mathcal
    {\widehat O}}\rangle}>1~,
\ee
where the absolute error is approximated by
\be{eq:absError}
\Delta\langle {\mathcal {\widehat O}}\rangle=\frac{ S[{\mathcal {\widehat O}}]}{\sqrt N}
\sqrt{\langle {\mathcal {\widehat O}}^2\rangle-\langle {\mathcal {\widehat O}}\rangle^2}\simeq 
\frac{  S[{\mathcal {\widehat O}}]}{\sqrt N }\sqrt{\langle {\mathcal {\widehat
      O}}^2\rangle}~.
\ee
The theoretical statistical significance 
is thus equal to the number of 
standard deviations to which the corresponding CP observable
can be determined to be non-zero.
For an ideal detector, a significance of, e.g., $S = 1$ implies
that the CP observables can be measured at the statistical 68\% confidence
level.
We remark that the theoretical statistical significance in Eq.~\rf{eq:EffAsy} 
is solely a theoretical definition.
Background and detector simulations for particle reconstruction
efficiencies are not included. In order to give realistic values 
of the statistical significances, a detailed Monte Carlo analysis 
would be required, which is however beyond the scope of the present work.
In the following we comment on the major SUSY and SM backgrounds
and discuss how they can be reduced.

\subsection{Lepton energy distributions and backgrounds \label{backgrounds}}

\begin{figure}[t]
\setlength{\unitlength}{1mm}
\begin{center}
\begin{picture}(130,70)
\put(-60,-205){\mbox{\epsfig{figure=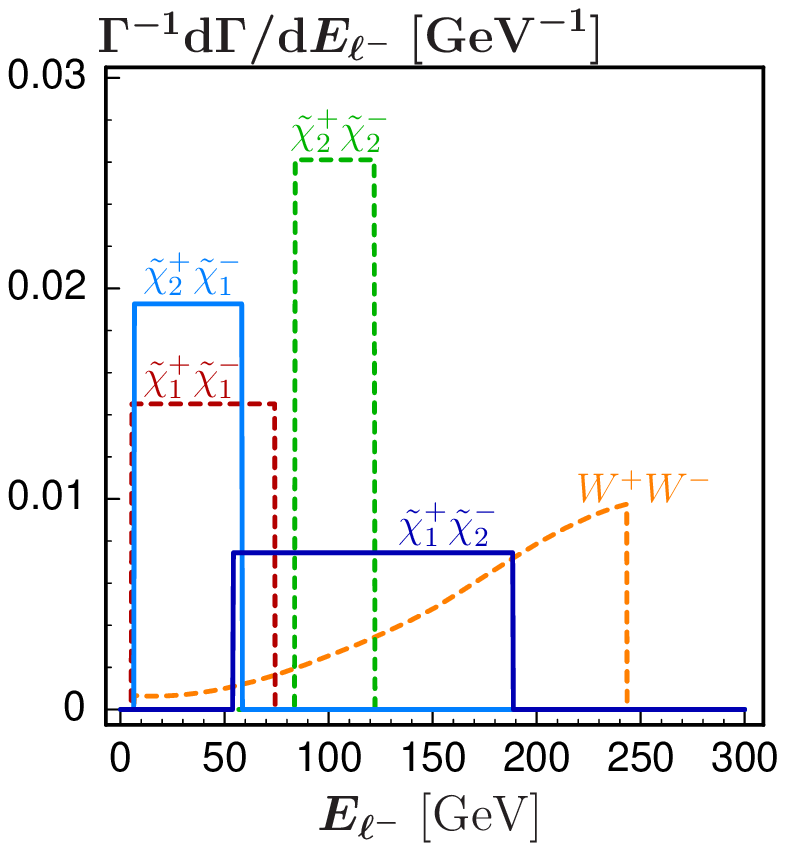,height=30cm,width=24cm}}}
\end{picture}
\end{center}
\vskip1cm
\caption{Normalized energy distributions of the signal leptons $\ell^-$ from 
chargino pair production $e^+e^-\to\tilde\chi_2^+\tilde\chi_1^-$;
$\tilde\chi_1^-\to\bar{\tilde\nu}_\ell~\ell^-$ (light blue line) and
$e^+e^-\to\tilde\chi_1^+\tilde\chi_2^-$;
$\tilde\chi_2^-\to \bar{\tilde\nu}_\ell~\ell^-$ (dark blue line), for the 
parameters as given in Table~\ref{tab1}.
Furthermore, we show the energy distributions of the background leptons $\ell^-$
from the reactions $e^+e^-\to W^+W^-$; $W^-\to\bar{\nu}_\ell~\ell^-$ 
(dashed orange line~\cite{Denner:1999dt}), $e^+e^-\to\tilde\chi_1^+\tilde\chi_1^-$; 
$\tilde\chi_1^-\to \bar{\tilde\nu}_\ell~\ell^-$ (dashed red line), and
$e^+e^-\to\tilde\chi_2^+\tilde\chi_2^-$; 
$\tilde\chi_2^-\to \bar{\tilde\nu}_\ell~\ell^-$ (dashed green line).}
\label{fig:fig2}
\end{figure}

A main SM background will be from $W$ pair production,
$e^+e^-\to W^+W^-$,
where both $W$'s decay leptonically
$W^+\to\nu_\ell~\ell^+$ and $W^-\to\bar{\nu}_\ell~\ell^-$. 
For $\sqrt s = 500$~GeV and unpolarized beams
the cross section is $\sigma(e^+e^-\to W^+W^-)=7.4$~pb~\cite{Beenakker:1994vn},
and $\sum_{\ell,\ell'}{\rm BR}(W^+ \to \nu_{\ell}~\ell^+)
{\rm BR}(W^- \to\bar{\nu}_{\ell'}~\ell'^-)=4.4\% $,
for $\ell,\ell' =e,\mu$~\cite{Yao:2006px}.
With a beam polarization of 
$({\mathcal P}_-,{\mathcal P}_+)=(-0.9,0.6)$,
the cross
section for $e^+e^-\to W^+W^-$ is about a factor $3$ larger~\cite{Beenakker:1994vn}.
A main SUSY background would be from the pair production
of equal charginos, 
$e^+e^-\to \tilde\chi^-_1\tilde\chi^+_1$ and
$e^+e^-\to \tilde\chi^-_2\tilde\chi^+_2$.
For the scenario as given in Table~\ref{tab1}, 
the cross sections are
$\sigma(e^+e^-\to \tilde\chi^-_1\tilde\chi^+_1)=974$~fb, and
$\sigma(e^+e^-\to \tilde\chi^-_2\tilde\chi^+_2)=145$~fb
at $\sqrt s = 500$~GeV with
$({\mathcal P}_-,{\mathcal P}_+)=(-0.9,0.6)$.
The signal cross section is 
$\sigma(e^+e^-\to \tilde\chi^+_1\tilde\chi^-_2)=527$~fb.
The chargino branching ratios are
${\rm BR}( \tilde\chi^+_1\to \tilde\nu_e~e^+)=33\%$ and
${\rm BR}( \tilde\chi^+_2\to \tilde\nu_e~e^+)=8\%$.

A large part of the background can be cut by
using the different energy distributions of the final state leptons.
In Fig.~\ref{fig:fig2}, we show the normalized energy distributions of 
the leptons $\ell^-$ 
stemming from the various reactions $e^+e^-\to\tilde\chi_i^+\tilde\chi_j^-$;
$\tilde\chi_j^-\to \tilde\nu_\ell~\ell^-$, where we have neglected the
effect of the longitudinal chargino polarization which may somewhat 
change the box-shape of the distributions.
For the energy distribution of the leptons $\ell^-$
from $e^+e^-\to W^+W^-$; $W^-\to\bar{\nu}_\ell~\ell^-$, 
we have fitted the curve in Fig.~3 of Ref.~\cite{Denner:1999dt}.

We can observe from Fig.~\ref{fig:fig2} that the 
energy distribution of the signal leptons $\ell^-$ from
$e^+e^-\to\tilde\chi_1^-\tilde\chi_2^+$, 
$\tilde\chi_1^-\to \bar{\tilde\nu}_\ell~\ell^-$
(light blue line), does only slightly overlap with that from 
the charge conjugated process $e^+e^-\to\tilde\chi_1^+\tilde\chi_2^-$,
$\tilde\chi_2^-\to \bar{\tilde\nu}_\ell~\ell^-$ (dark blue line) in this scenario.
In the first place, this is essential for a measurement of the CP observables,
since they change sign for the charge conjugated process.
Secondly, the background from equal charginos pair production 
$e^+e^-\to \tilde\chi^-_2\tilde\chi^+_2$ can be totally eliminated, since both 
leptons have an energy in the interval $E_\ell \in [84,122]~{\rm GeV}$.
As discussed above, a signal event has
a lepton with energy $E_\ell \in [7,59]~{\rm GeV}$,
and an oppositely charged lepton with energy $E_\ell \in [54,189]~{\rm GeV}$.
On the other hand, the lepton energy distribution from 
$\tilde\chi^+_1\tilde\chi^-_1$ production overlaps 
with that of a signal lepton (the one from the $\tilde\chi^\pm_2$ decay) 
only in the interval $E_\ell \in [54,74]~{\rm GeV}$.
The background from $\tilde\chi^+_1\tilde\chi^-_1$ production
can be eliminated, if we require one lepton with $E_\ell>74$~GeV.

The energy distribution of the leptons from $W^\pm$ decays overlaps with that
from the signal in the whole range, see Fig.~\ref{fig:fig2}.
However, the signal involves one lepton with a rather low energy 
$E_\ell \in [7,59]~{\rm GeV}$, where the background is low.
In this interval, the signal is about twice as large as the $W^\pm$ background.
Thus, as any CP-even background, which contributes only to the denominator
but not to the numerator of the CP observables,
it will reduce them and the corresponding significances only slightly.
Moreover, the high energetic background leptons from $W^\pm$ decays can be 
significantly reduced by the cut $E_\ell \lsim 189$~GeV,
see Fig.~\ref{fig:fig2}.

\medskip

Additional important SUSY background originates from the following
processes~\cite{Kalinowski:2008sa,Kato:2004bg}:
(a)~stau pair production $e^+e^-\to \tilde\tau^+\tilde\tau^-$, with subsequent decays
$\tilde\tau\to\tau\tilde\chi^0_1$, followed by leptonic tau decays,
(b)~selectron and smuon production $e^+e^-\to\tilde \ell^+\tilde \ell^-$, 
$\ell = e,\mu$,  followed by $\tilde\ell\to\ell\tilde\chi^0_1$, and 
(c)~neutralino production $e^+e^-\to\tilde\chi_i^0\tilde\chi_j^0$ 
and their decays into leptons via sleptons.

Additional SM background reactions are~\cite{Kalinowski:2008sa,Kato:2004bg}:
(i)~pho\-ton-induced tau pair production $e^+e^-\to e^+e^-\tau^+\tau^-$,
followed by leptonic tau decays,
(ii)~pho\-ton-induced $W$ pair production $e^+e^-\to e^+e^-W^+W^-$, with
$W^\pm\to \ell^\pm\nu_\ell$, 
(iii)~tau pair production followed by leptonic tau decays,
(iv)~$Z$ pair production followed by the decays into leptons, and
(v)~single boson production, $e^+e^-Z$ and $\nu\bar\nu Z$ with 
$Z\to\ell^+\ell^-,\nu\bar\nu$. 

Recently a detailed NLO study of chargino pair production
$e^+e^-\to\tilde\chi^+_1\tilde\chi^-_1$
and decay $\tilde\chi^\pm_1\to \ell^\pm \tilde\nu_\ell$
at a linear collider with $\sqrt{s}=1$~TeV has been performed~\cite{Kalinowski:2008sa}. 
The SM and SUSY backgrounds to the experimental signature
$\e^\pm\mu^\mp+/\!\!\!\!E$ have been taken into account.
A signal to background ratio of $0.62$ has been obtained
after appropriate cuts~\cite{Kalinowski:2008sa}.
In particular, the background from photon-induced $\tau$ pair production, 
which may exceed the size of the signal cross section by a factor of $10^4$, 
has been reduced by a factor of $10^6$.
Further the authors of Ref.~\cite{Kalinowski:2008sa} have shown 
that SM processes lead to a flat background distribution,
which can be easily subtracted, while SUSY backgrounds are more challenging,
since their kinematic distributions are similar to the signal in general.
We expect that our proposed discrimination criteria for the lepton energies,
together with cuts as applied in~\cite{Kalinowski:2008sa,Kato:2004bg},
will enhance the signal to background ratio also for our observables.

\section{Numerical results \label{numerics}}

We present numerical results for the 
CP-sensitive observables defined in Section~\ref{observables} 
for chargino production and decay,
$e^+e^-\to \CH^\pm_1\CH^\mp_2 \to 
\bar{\tilde\nu}_\ell ~\ell^+~\tilde\nu_{\ell'} ~\ell'^-$, 
for $\ell,\ell' = e, \mu$.
We study the dependence of the CP observables on the phase 
of the higgsino mass parameter $\mu=|\mu|e^{i\phi_\mu}$ 
in the framework of the general MSSM, where
restrictions on $\phi_\mu$ from the electron and neutron
EDMs are less severe compared to the constrained MSSM~\cite{EDMs,EDMs1}.
For example, the cancellation of various contributions to the EDMs allow
for the possibility of $\phi_\mu\sim {\mathcal O}(1)$~\cite{EDMs1}. 
Since our analysis does not include all relevant parameters
necessary to predict the actual values of the EDMs, we do not
take the EDMs into account, and show the full
$\phi_\mu$-dependence of the observables.

Our study is for the ILC 
with $\sqrt{s}=500$~GeV and longitudinal 
beam polarizations 
$({\mathcal P}_-,{\mathcal P}_+)=(-0.9,0.6)$.
This choice enhances the $\tilde{\nu}_e$
exchange contribution, yielding
larger cross sections and CP observables. We also provide numerical 
results for the chargino cross sections and the branching ratios. Furthermore, we
give the theoretical statistical significances to which the CP-sensitive 
observables can be determined to be non-zero.

For the calculation of the chargino decay widths and branching ratios
we consider their two-body decays~\cite{Kittel:2004rp}
\baq{eq:BRs}
\CH^\pm_{1,2} &\to& W^\pm~\CH^0_n,\, e^\pm~\tilde{\nu}_e,
\,\mu^\pm~\tilde{\nu}_\mu,\, \tau^\pm~\tilde{\nu}_\tau,
\,\nu_e~\tilde{e}^\pm_L,\,\nu_\mu~\tilde{\mu}^\pm_L,
\,\nu_\tau~\tilde{\tau}^\pm_{1,2}~,\nonumber\\
\CH^\pm_{2} &\to& Z~\CH^\pm_1,\,h~\CH^\pm_1~.
\eaq
We assume the GUT inspired relation $|M_1|=5/3 \, M_2 \tan^2 \Theta_W$,
and in the stau sector we fix the trilinear scalar
coupling parameter $A_\tau = 250$~GeV.

\begin{table}[h]
\vspace{1cm}
\begin{center}
\begin{tabular}{ccccccc} \hline
 
  \multicolumn{1}{c}{$M_2$} 
& \multicolumn{1}{c}{$|\mu|$}
& \multicolumn{1}{c}{$\phi_\mu$}  
& \multicolumn{1}{c}{$\tan{\beta}$}  
& \multicolumn{1}{c}{$m_{\tilde{\nu}_{e,\mu}}$}
& \multicolumn{1}{c}{$m_{\chi_1}$}
& \multicolumn{1}{c}{$m_{\chi_2}$}
\\\hline
 
  \multicolumn{1}{c}{152} 
& \multicolumn{1}{c}{200}
& \multicolumn{1}{c}{$0.5\pi$}  
& \multicolumn{1}{c}{3} 
& \multicolumn{1}{c}{103}
& \multicolumn{1}{c}{125}
& \multicolumn{1}{c}{246}
\\\hline
\end{tabular}
\\[5.0ex]
\caption{Input parameters $M_2$, $|\mu|$, $\phi_\mu$, $\tan\beta$, 
and $m_{\tilde{\nu}_{e,\mu}}$.
All mass parameters are given in GeV. 
The other masses are $m_{\tilde e_R}=107$~GeV, $m_{\tilde e_L}=125$~GeV, 
$m_{\tilde\tau_1}=106$~GeV, $m_{\tilde\tau_2}=127$~GeV, $m_{\chi^0_1}=73$~GeV,
$m_{\chi^0_2}=127$~GeV, $m_{\chi^0_3}=208$~GeV, $m_{\chi^0_4}=247$~GeV.
\label{tab1}}
\end{center}
\end{table}

Before resuming with the numerical investigation, we
address the parameter dependence of the CP-sensitive coupling factors
${\rm Im}(O_{12}^{'L} O_{12}^{'R*})$ and ${\rm Im}(V_{11}^* V_{21}
O_{12}^{'R})$ on which our CP-sensitive observables depend, see
Eqs.~\rf{eq:Toddpart1} and~\rf{eq:Toddpart2}.
When we expand them by using the parametrization of the chargino mixing matrices
$U$ and $V$, we find
\baq{eq:expansion}
{\rm Im}(V_{11}^* V_{21} O_{12}^{'R})=2~{\rm Im}(O_{12}^{'L} O_{12}^{'R*})= 
\frac{1}{8}\sin2\theta_1\sin2\theta_2
\sin(\phi_1-\phi_2+\gamma_1-\gamma_2)~,
\nonumber\\
\eaq
with the chargino mixing angles $\theta_1, \theta_2$, and 
the phases $\phi_1, \gamma_1,\phi_2,\gamma_2$ of the matrices $U,V$.
Their explicit dependence on the parameters of the chargino system
can be found in \cite{staplex,Choi:1998ut}.
In particular, the phases are zero when $\phi_\mu$ is zero.
One finds that the CP-sensitive coupling factors (and therefore the CP observables) are
largest for large gaugino-higgsino mixing, i.e. for $M_2\sim |\mu|$.
Furthermore, a small value for $\tan\beta$ is preferable, as
$\tan\beta\to \infty$ results in
$\phi_1,\gamma_2\to \phi_\mu$ and $\phi_2,\gamma_1\to 0$, leading to
$\sin(\phi_1-\phi_2+\gamma_1-\gamma_2)\to 0$.
We therefore choose a mixed scenario with small $\tan\beta$, see the 
parameters in Table~\ref{tab1}.
   
\begin{figure}[t]
\setlength{\unitlength}{1mm}
\begin{center}
\begin{picture}(130,70)
\put(-75,-205){\mbox{\epsfig{figure=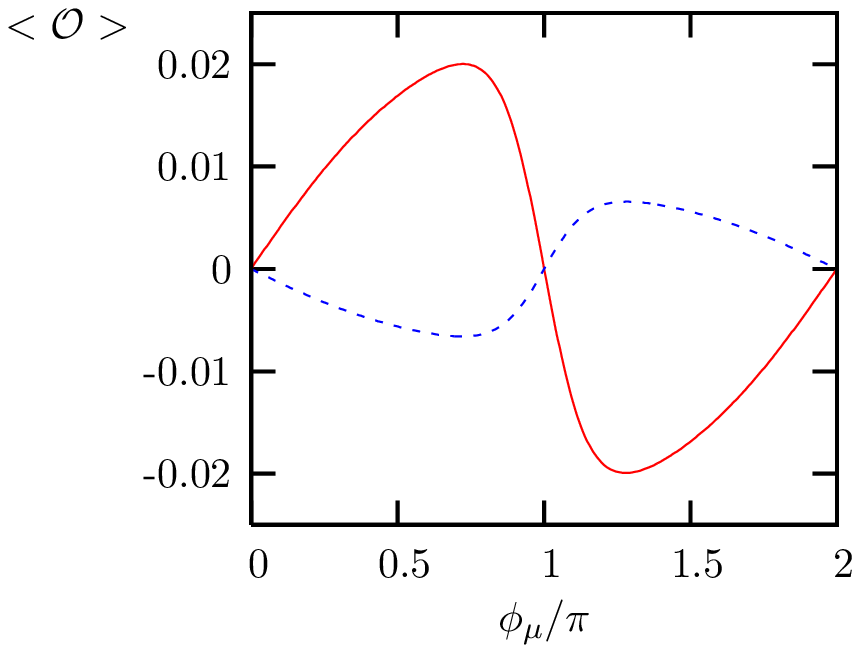,height=30cm,width=18cm}}}
\put(5,-205){\mbox{\epsfig{figure=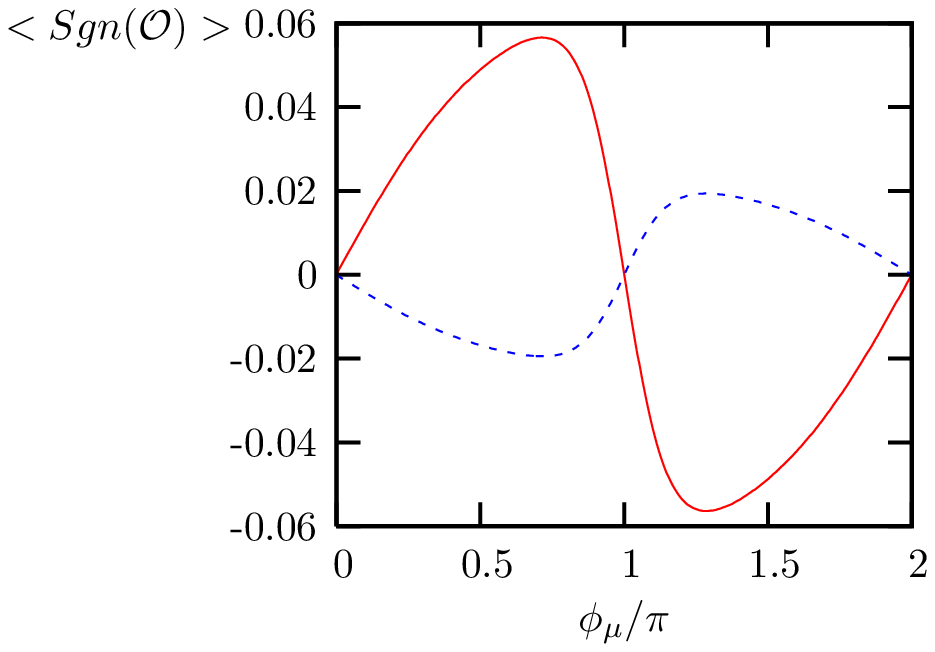,height=30cm,width=18cm}}}
\put(5,52){\mbox{{\bf (a)}}}
\put(23,50){\mbox{\footnotesize{$\langle {\mathcal O}_T^{\rm prod}\rangle$}}}
\put(9,22){\mbox{\footnotesize{$\langle {\mathcal O}_T \rangle$}}}
\put(90,52){\mbox{{\bf (b)}}}
\put(110,50){\mbox{\footnotesize{$\langle {\rm Sgn}({\mathcal O}_T^{\rm prod})\rangle$}}}
\put(91,19.5){\mbox{\footnotesize{$\langle {\rm Sgn}({\mathcal O}_T)\rangle$}}}
\end{picture}
\end{center}
\vskip1cm
\caption{$\phi_\mu$-dependence of the CP-sensitive observables (a) 
$\langle {\mathcal O}_T^{\rm prod}\rangle$ (red solid line) and
$\langle {\mathcal O}_T\rangle$ (blue dashed line), 
and (b) 
$\langle {\rm Sgn}({\mathcal O}_T^{\rm prod})\rangle$ (red solid line) and
$\langle {\rm Sgn}({\mathcal O}_T)\rangle$ (blue dashed line), 
{for chargino production and decay}
$e^+e^-\to \CH^+_1\CH^-_2 \to \tilde\nu_e ~e^+~\bar{\tilde\nu}_e ~e^-$,
for the scenario defined in Table~\ref{tab1},
at $\sqrt s=500$~GeV with longitudinal beam polarizations 
$({\mathcal P}_-,{\mathcal P}_+)=(-0.9,0.6)$.
}
\label{fig:fig3}
\end{figure}

In Fig.~\ref{fig:fig3}a, we show the $\phi_\mu$-dependence of 
the CP-sensitive observables
$\langle {\mathcal O}_T\rangle$ and 
$\langle {\mathcal O}_T^{\rm prod}\rangle$, 
Eq.~\rf{eq:Obs} and~\rf{eq:ObsProd}, respectively.
One can clearly see their asymmetric dependence on
$\phi_\mu$. At $\phi_\mu=0 \,(mod \,\pi)$, the chargino couplings are real and therefore
the CP-sensitive observables vanish. The observables attain their
largest values at $\phi_\mu=0.75 \pi$ of about 
$\langle {\mathcal O}_T\rangle=-6.5 \cdot 10^{-3}$ and 
$\langle {\mathcal O}_T^{\rm prod}\rangle=2 \cdot 10^{-2}$. 
Fig.~\ref{fig:fig3}b shows the $\phi_\mu$-dependence of 
the CP-sensitive observables
$\langle {\rm Sgn}({\mathcal O}_T)\rangle$  and
$\langle {\rm Sgn}({\mathcal O}_T^{\rm prod})\rangle$.
At $\phi_\mu=0.75 \pi$ the observables 
reach $\langle {\rm Sgn}({\mathcal O}_T)\rangle=-1.9\%$ and 
$\langle {\rm Sgn}({\mathcal O}_T^{\rm prod})\rangle= 5.6\%$.

\begin{figure}[t]
\setlength{\unitlength}{1mm}
\begin{center}
\begin{picture}(130,70)
\put(-80,-205){\mbox{\epsfig{figure=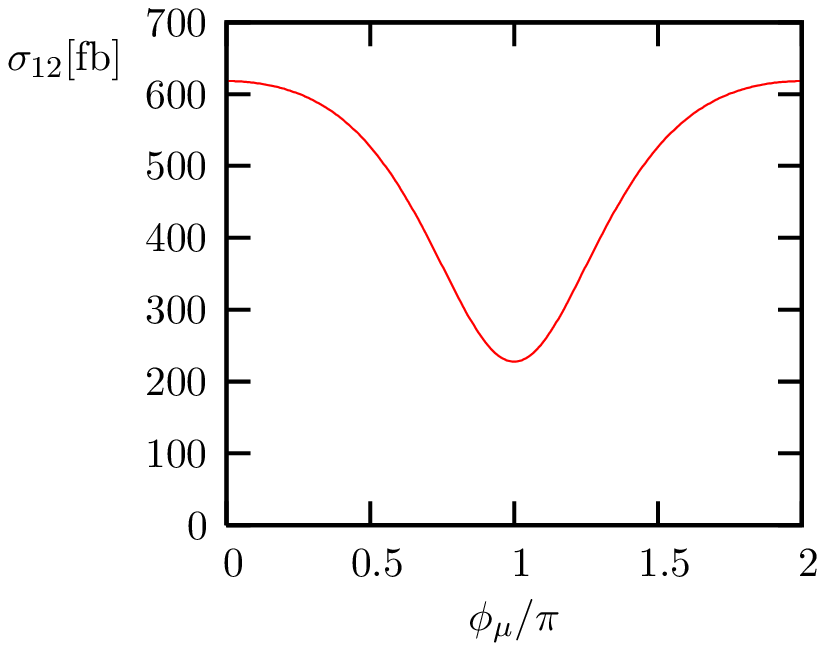,height=30cm,width=20cm}}}
\put(5,-205){\mbox{\epsfig{figure=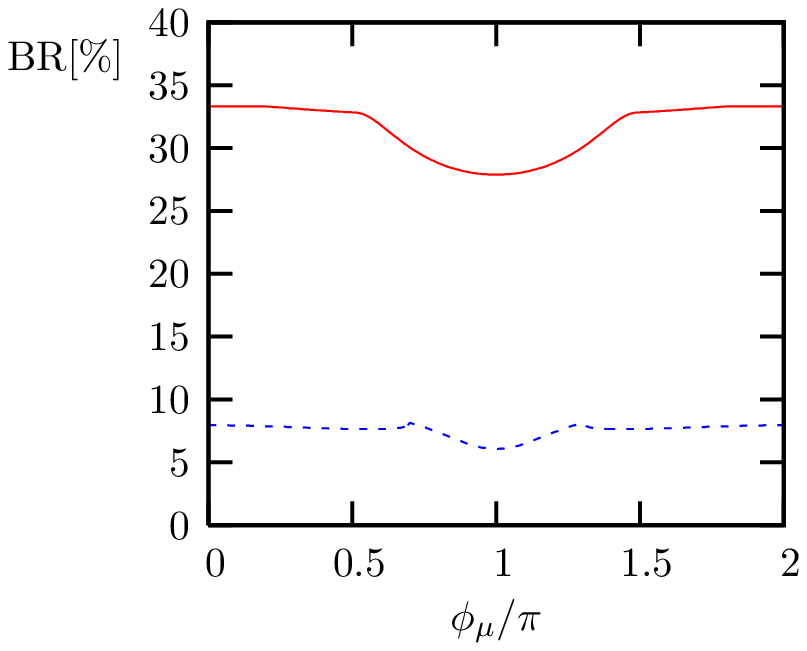,height=30cm,width=20cm}}}
\put(-73.5,-275){\mbox{\epsfig{figure=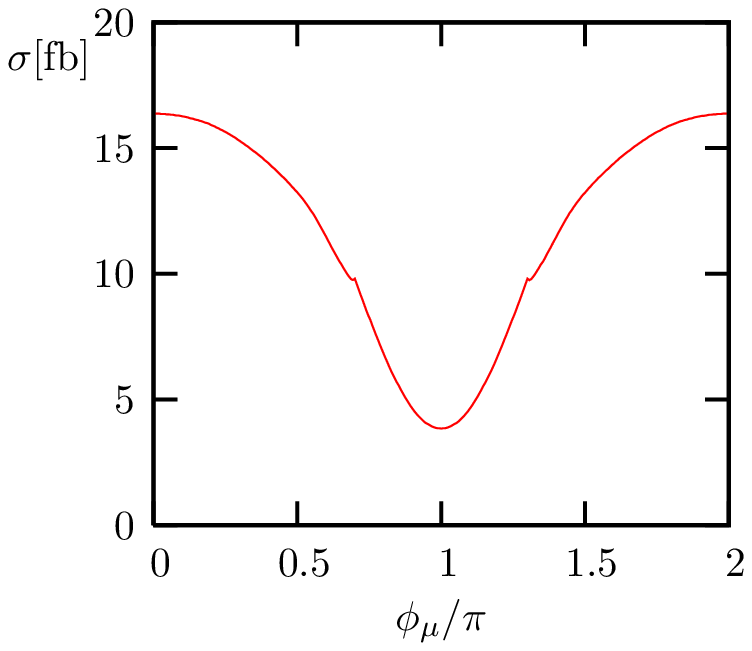,height=30cm,width=20cm}}}
\put(6.5,-275){\mbox{\epsfig{figure=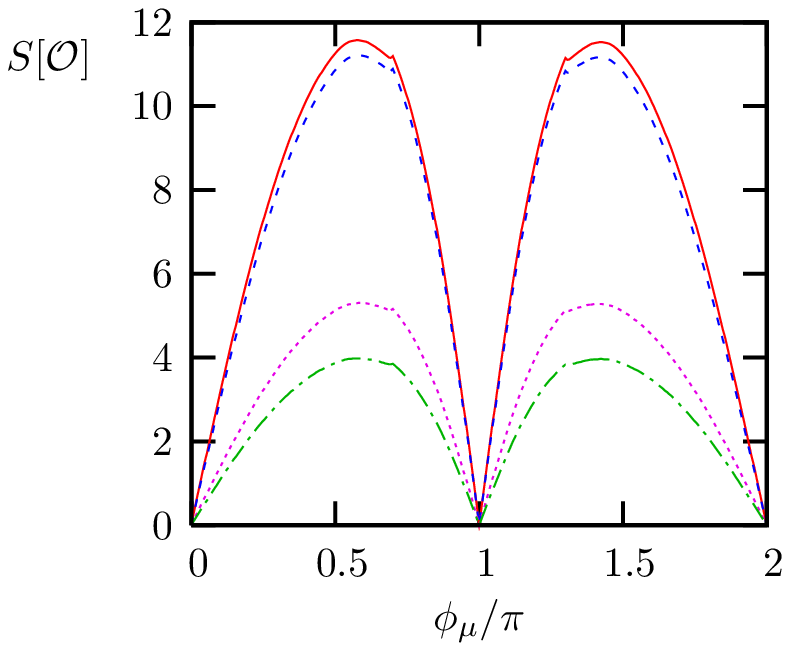,height=30cm,width=20cm}}}
\put(45,54){\mbox{{\bf (a)}}}
\put(128,54){\mbox{{\bf (b)}}}
\put(97,38){\mbox{\footnotesize{BR$(\tilde\chi^+_1\to \tilde\nu_e~e^+)$}}}
\put(45,-17){\mbox{{\bf (c)}}}
\put(128,-17){\mbox{{\bf (d)}}}
\put(97,20){\mbox{\footnotesize{BR$(\tilde\chi^-_2\to \bar{\tilde\nu}_e~e^-)$}}}
\end{picture}
\end{center}
\vskip8cm
\caption{$\phi_\mu$-dependence of (a) the chargino production cross 
section $\sigma_{12}\equiv\sigma(e^+e^-\to \CH^+_1\CH^-_2)$, 
(b) the branching ratios BR$(\CH^+_1\to  \tilde\nu_e ~e^+)$ (red solid line)
and BR$(\CH^-_2\to  \bar{\tilde\nu}_e ~e^-)$ (blue dashed line), (c) the cross section
$\sigma\equiv\sigma(e^+e^-\to \CH^+_1\CH^-_2 \to \tilde\nu_e e^+
\bar{\tilde\nu}_e e^-)$, and (d) the theoretical statistical significances 
$S[{\rm Sgn}({\mathcal O}_T^{\rm prod})]$ (red solid line),
$S[{\mathcal O}_T^{\rm prod}]$ (blue dashed line),
$S[{\mathcal O}_T]$ (magenta dotted line),
and $S[{\rm Sgn}({\mathcal O}_T)]$ (green dot-dashed line), 
for the scenario defined in Table~\ref{tab1},
at $\sqrt s=500$~GeV with longitudinal beam polarizations 
$({\mathcal P}_-,{\mathcal P}_+)=(-0.9,0.6)$,
and for (d), with an integrated luminosity ${\mathcal L}=500~{\rm fb}^{-1}$.}
\label{fig:fig4}
\end{figure}

In Fig.~\ref{fig:fig4}a, we show the chargino production cross
section $\sigma_{12}\equiv\sigma(e^+e^-\to \CH^+_1\CH^-_2)$ 
as a function of $\phi_{\mu}$. 
The production cross section varies between $\sigma_{12}=618$~fb for 
$\phi_{\mu}=0 \,(mod \, 2\pi)$, and $\sigma_{12}=227$~fb for $\phi_{\mu}=\pi$.
Fig.~\ref{fig:fig4}b
shows the branching ratios BR$(\CH^+_1\to  \bar{\tilde\nu}_e ~e^+)$
and BR$(\CH^-_2\to  \bar{\tilde\nu}_e ~e^-)$, which are about $32\%$ and
$7\%$, respectively.
The $\phi_\mu$-dependence of the cross section 
$\sigma\equiv\sigma(e^+e^-\to \CH^+_1\CH^-_2 \to \tilde\nu_e~e^+~\tilde{\bar{\nu}}_e~e^-)$
is given in Fig.~\ref{fig:fig4}c. 
In contrast to the CP-sensitive observables, 
the cross section shows a symmetric dependence on $\phi_\mu$. 
The phase ambiguity $\phi_\mu\leftrightarrow2\pi-\phi_\mu$
of the higgsino mass parameter $\mu$ can only be resolved by
a measurement of CP-sensitive observables.
In Fig.~\ref{fig:fig4}d, we present the theoretical statistical significances,
$S[{\mathcal {\widehat O}}]$, as defined in Eq.~\rf{eq:EffAsy},
for an integrated luminosity ${\mathcal L}=500$~fb$^{-1}$.
For $\langle {\rm Sgn}({\mathcal O}_T)\rangle$ and 
$\langle {\mathcal O}_T\rangle$, which do not require 
a reconstruction of the production plane, the theoretical statistical significances
reach $4$ and $5$ standard deviations, respectively. 
The theoretical statistical
significances of the CP-sensitive observables 
$\langle {\rm Sgn}({\mathcal O}_T^{\rm prod})\rangle$ and 
$\langle {\mathcal O}_T^{\rm prod}\rangle$ are at the $12$--$\sigma$ level
for the considered scenario.

We note that also other leptonic chargino decay chains, i.e.  
$\CH^\pm_{1,2}\to\tilde\ell^\pm~\nu; \tilde\ell^\pm\to\tilde\chi^0_1~\ell^\pm$ and 
$\CH^\pm_{1,2}\to W^\pm\tilde\chi^0_1; W^\pm\to \nu~\ell^\pm$, can be used for
measuring the spin-spin correlation terms of the chargino production
amplitude. The inclusion of these decay chains can be done in a 
similar fashion as in Ref.~\cite{CPsbottom}. 
However, we expect that this would not
lead to larger statistical significances of the CP-sensitive observables.
Although the total number of events is increased, the 
CP-sensitive observables are reduced, since they have
to be weighted with the corresponding decay branching ratios~\cite{CPsbottom}.

Finally, note that the T-odd products, 
Eqs.~\rf{eq:Todd1} and~\rf{eq:Todd1prod}, can also be used for the
definition of CP-sensitive observables in neutralino production
with subsequent decays. This is possible for the pair production
$\sigma(e^+e^-\to\tilde\chi^0_i\tilde\chi^0_j)$, with
$i> j\geq2$. The neutralino polarizations can then again be obtained 
from the lepton distributions, e.g., in the decays 
$\tilde\chi^0_{i,j}\to \tilde\ell^\pm~\ell^\mp$~\cite{MarkDiplom}.

\clearpage

\section{Summary and conclusions \label{conclusion}}

We have proposed novel CP-sensitive observables 
in chargino production $e^+e^-\to \CH^\pm_1\CH^\mp_2$.
These CP observables are sensitive to the phase of the
higgsino parameter $\mu$. They arise on tree-level, and rely on T-odd products
in the chargino spin-spin correlations. These are the terms of the 
matrix element, which include the polarizations of both charginos,
with one component perpendicular to the production plane. 
The chargino polarization can be deduced from the distributions
of their leptonic decay products
$\CH^\pm_{1,2}\to \tilde\nu_\ell~\ell^\pm$, $\ell=e,\mu$.

In order to probe the CP-sensitive spin-spin correlation terms,
we have identified two different T-odd products.
The first one, ${\mathcal O}_T$, does not involve the
chargino momentum, which has the
advantage that it is not necessary to reconstruct the production plane. 
We recall that other T-odd products 
proposed in the literature always require such a 
reconstruction, if only one leptonic chargino decay is considered. 
The second T-odd product, ${\mathcal O}^{\rm prod}_T$, in
contrast includes the chargino momentum.
Based on these T-odd products, we have defined two sorts of CP-sensitive
observables. One is an up-down asymmetry, giving the 
difference of events with positive and negative T-odd products.
The other sort of CP-sensitive observables are the expectation 
values of the T-odd products for the event sample.

In the numerical study, we have found that the observables are
largest in mixed scenarios with small $\tan\beta$.
We have defined theoretical significances 
to decide, which CP observable is most sensitive to the
CP phase $\phi_\mu$.
For a linear collider with $\sqrt s=500$~GeV and 
longitudinally polarized beams,
$({\mathcal P}_-,{\mathcal P}_+)=(-0.9,0.6)$,
with an integrated luminosity of ${\mathcal L}=500~{\rm fb}^{-1}$,
the CP-sensitive observables that are based on the T-odd product ${\mathcal O}_T$ 
yield $S[{\mathcal O}_T]\lsim 5$.
We find larger significances $S[{\mathcal O}^{\rm prod}_T]\lsim 12$
for the CP-sensitive observables that are
based on ${\mathcal O}^{\rm prod}_T$.
Thus the largest CP-violating effects are obtained
if the chargino production plane can be reconstructed.
However, only a detailed experimental study with
background and detector simulations 
can show whether the CP-sensitive observables 
are accessible.

Finally, we remark that our proposed method 
for analyzing T-odd products in the spin-spin correlations terms
can also be used for the definition of CP-sensitive observables in 
other fermion pair production processes, such as neutralino productions.

\section*{Acknowledgements}

We thank K. Desch and P. Wienemann for very helpful discussions
on the background from $W$ pairs. M.T. thanks R. K\"ogerler for very 
useful discussions and encouragement.  
This work is supported by the 'Fonds zur F\"orderung der
wissenschaftlichen Forschung' (FWF) of Austria, project. No. P18959-N16.
The authors acknowledge support from EU under the MTRN-CT-2006-035505 
and MTRN-CT-2006-503369 network programs.
A. B. was supported by the Spanish grants SAB 2006-0072, FPA 2005-01269
and FPA 2005-25348-E of Ministero de Educacion y Ciencia.
O.K. was supported by the SFB Transregio
33: The Dark Universe.

\begin{appendix}
\setcounter{equation}{0}
\renewcommand{\thesubsection}{\Alph{section}.\arabic{subsection}}
\renewcommand{\theequation}{\Alph{section}.\arabic{equation}}

\section*{Appendix}

\section{Quantities $D$ and $\Sigma^{c}_{D}$ in the spin density matrix
formalism \label{Decaydensity}}

The coefficients in Eq.~\rf{eq:amplitudesquared} of the chargino 
decay matrices for $\CH_k^+\to\ti\nu_\ell ~\ell^+$
and $\CH_k^-\to\bar{\ti\nu}_{\ell}~\ell^-$, $\ell=e,\mu$, are
\be{eq:Dlep}
D_k=\frac{g^2}{2} |V_{k1}|^2 (m^2_{\chi_k}-m^2_{\ti\nu_{\ell}})~,
\ee
and
\be{eq:Slep}
\Sigma^c_{D_k}=\pm g^2 |V_{k1}|^2 m_{\chi_k} 
(s_{\chi^\mp_k}^c  p_{\ell^\mp})~.
\ee
The positive sign in Eq.~\rf{eq:Slep} holds for the decay 
$\CH_k^-\to\bar{\ti\nu}_\ell~ \ell^-$,  and the negative sign 
for the charge conjugated decay $\CH_k^+\to\ti\nu_\ell~\ell^+$.

\section{Momentum and polarization vectors\label{Vectors}}

We choose a coordinate system with the $z$-axis
along the $\vec{p}_{e^-}$ direction in the center-of-mass system.
The 4-momenta of the charginos $\ti\chi^-_i$ and $\ti\chi^+_j$ are 
\baq{eq:momentumchar}
p_{\chi^+_i}&=&q
(E_{\chi_i}/q,-\sin\theta,0,-\cos\theta)~,\nonumber\\[2mm]
p_{\chi^-_j}&=&q
(E_{\chi_j}/q,\sin\theta,0,\cos\theta)~,
\eaq
with their energies and common momentum
\be{eq:energy}
E_{\chi_{i,j}}=\frac{s+m^2_{\chi_{i,j}}-m^2_{\chi_{j,i}}}{2 \sqrt{s}}~,\qquad
q=\frac{\lambda^{\frac{1}{2}}(s,m^2_{\chi_i},m^2_{\chi_j})}{2 \sqrt{s}}~,
\ee
respectively, with
the scattering angle $\theta \angle (\vec{p}_{e^-},\vec{p}_{\chi^-_j})$, 
and the kinematic function $\lambda(a,b,c)=a^2+b^2+c^2-2(a b + a c + b c)$.
The azimuthal angle can be set to zero, due to rotational invariance
around the beam axis.

The three spin basis vectors of $\ti\chi^+_k$ and $\ti\chi^-_k$
are chosen to be 
\baq{eq:polvec}
s^1_{\chi^\pm_k}&=&\left(0,\frac{\vec{s}_{\chi^\pm_k}^{\,2}\times\vec{s}_{\chi^\pm_k}^{\,3}}
{|\vec{s}_{\chi^\pm_k}^{\,2}\times\vec{s}_{\chi^\pm_k}^{\,3}|}\right)=
\pm(0, \cos\theta,0,-\sin\theta)~,
\nonumber \\[3mm]
s^2_{\chi^\pm_k}&=&\left(0,
\frac{\vec{p}_{e^-}\times \vec{p}_{\chi^-_k}}
{|\vec{p}_{e^-}\times\vec{p}_{\chi^-_k}|}\right)=
(0,0,1,0)~,
\nonumber \\[3mm]
s^3_{\chi^\pm_k}&=&\frac{1}{m_{\chi_k}}
\left(q, 
\frac{E_{\chi_k}}{q}~\vec{p}_{\chi^\pm_k} \right)=
\frac{E_{\chi_k}}{m_{\chi_k}}
(q/E_{\chi_k},\mp \sin\theta,0,\mp\cos\theta)~.
\eaq
They fulfill the orthonormality relations
$s_{\chi_k^\pm}^{c}\cdot s_{\chi_k^\pm}^{d}=-\delta^{cd}$ and
$s_{\chi_k^\pm}^{c}\cdot p_{\chi_k^\pm}=0$.
The 4-momenta of the leptons in the decays $\ti\chi^+_i
\to\ti\nu_\ell~\ell^+$ and $\ti\chi^-_j \to\bar{\ti\nu}_{\ell'}~\ell'^-$ are 
\be{eq:fourlp}
p_{\ell^+}=
|\vec{p}_{\ell^+}| (1,\cos\phi_{\ell^+}
\sin\theta_{\ell^+},
\sin\phi_{\ell^+} \sin\theta_{\ell^+},
\cos\theta_{\ell^+})~,
\ee
\be{eq:fourlm}
p_{\ell'^-}=
|\vec{p}_{\ell'^-}| (1,\cos\phi_{\ell'^-}
\sin\theta_{\ell'^-},
\sin\phi_{\ell'^-} \sin\theta_{\ell'^-},
\cos\theta_{\ell'^-})~,
\ee
respectively, with
\be{eq:lepmom}
|\vec{p}_{\ell^+}|=\frac{m^2_{\chi_i}-
m^2_{\ti\nu_\ell}}{2(E_{\chi_i}+q\cos\vartheta_{\ell^+})}~,
\quad
|\vec{p}_{\ell'^-}|=\frac{m^2_{\chi_j}-
m^2_{\ti\nu_{\ell'}}}{2(E_{\chi_j}-q\cos\vartheta_{\ell'^-})}~,
\ee
and
\baq{eq:angle}
\cos\vartheta_{\ell^+}&=&\sin\theta \sin\theta_{\ell^+} \cos\phi_{\ell^+}+
\cos\theta \cos\theta_{\ell^+}~,
\nonumber\\[2mm]
\cos\vartheta_{\ell'^-}&=&\sin\theta \sin\theta_{\ell'^-} \cos\phi_{\ell'^-}+
\cos\theta \cos\theta_{\ell'^-}~.
\eaq
%

\section{Spin-spin correlation terms in the laboratory system\label{Spin-spin-terms}}

The spin-spin correlation terms  in the laboratory system are
\begin{eqnarray}
\Sigma^{12}_P(Z Z)
&=&-\frac{g^4}{2\cos^4\Theta_W}|\Delta(Z)|^2 (L^2_e c_{-+}+R^2_e c_{+-})
{\rm Im}(O_{ij}^{'L} O_{ij}^{'R*})~E_{\chi_i} s q \sin^2\theta,
\label{eq:ZZ12}\\[2mm]
\Sigma^{12}_P(Z \ti\nu_e)
&=&\frac{g^4}{4\cos^2\Theta_W}
\Delta(Z)\Delta(\ti\nu_e)^*L_e~c_{-+}~
{\rm Im}(V_{i1}^* V_{j1} O_{ij}^{'R})~E_{\chi_i} s q \sin^2\theta~,
\label{eq:Zsnu12}\\[2mm]
\Sigma^{23}_P(Z Z)
&=&\frac{g^4}{4\cos^4\Theta_W}|\Delta(Z)|^2 (L^2_e c_{-+}+R^2_e c_{+-})
{\rm Im}(O_{ij}^{'L} O_{ij}^{'R*})~m_{\chi_j} s q \sin2\theta~,
\label{eq:ZZ23}\\[2mm]
\Sigma^{23}_P(Z \ti\nu_e)
&=&-\frac{g^4}{8\cos^2\Theta_W}
\Delta(Z)\Delta(\ti\nu_e)^*L_e~c_{-+}~
{\rm Im}(V_{i1}^* V_{j1} O_{ij}^{'R})~m_{\chi_j} s q \sin2\theta~,
\label{eq:Zsnu23}
\end{eqnarray}
which we obtain by inserting the momenta and spin vectors
in the laboratory system, Eqs.~\rf{eq:momentumchar} and~\rf{eq:polvec},
into Eqs.~\rf{eq:Toddpart1} and~\rf{eq:Toddpart2}. 
In order to obtain the terms $\Sigma^{21}_P(Z Z)$ and $\Sigma^{21}_P(Z\ti\nu_e )$,
one has to exchange  $E_{\chi_i}\to -E_{\chi_j}$ in 
Eqs.~(\ref{eq:ZZ12}) and~(\ref{eq:Zsnu12}).
In order to obtain the terms $\Sigma^{32}_P(Z Z)$ and $\Sigma^{32}_P(Z\ti\nu_e )$,
one has to exchange $m_{\chi_j} \to - m_{\chi_i}$ in 
Eqs.~(\ref{eq:ZZ23}) and~(\ref{eq:Zsnu23}).

\end{appendix}

\end{document}